\documentclass[aps,prl,twocolumn,floatfix,superscriptaddress]{revtex4-1}

\usepackage{graphicx}
\usepackage[colorlinks=true,linkcolor = blue,
            urlcolor  = blue,
            citecolor = blue,
            anchorcolor = black]{hyperref}
\usepackage{siunitx}
\usepackage{amssymb}

\begin{document}
\title{Superconductivity in type-II Weyl-semimetal $\mathrm{WTe_2}$ induced by a normal metal contact}

\author{Artem Kononov}
\email{Artem.Kononov@unibas.ch}
\affiliation{\footnotesize Department of Physics, University of Basel, Klingelbergstrasse 82, CH-4056 Basel, Switzerland}

\author{Martin Endres}
\affiliation{\footnotesize Department of Physics, University of Basel, Klingelbergstrasse 82, CH-4056 Basel, Switzerland}

\author{Gulibusitan Abulizi}
\affiliation{\footnotesize Department of Physics, University of Basel, Klingelbergstrasse 82, CH-4056 Basel, Switzerland}

\author{Kejian Qu}
\affiliation{\footnotesize Department of Materials Science and Engineering, University of Tennessee, Knoxville, TN 37996, United States}

\author{Jiaqiang Yan}
\affiliation{\footnotesize Materials Science and Technology Division, Oak Ridge National Laboratory, Oak Ridge, TN 37831, United States}
\affiliation{\footnotesize Department of Materials Science and Engineering, University of Tennessee, Knoxville, TN 37996, United States}

\author{David G. Mandrus}
\affiliation{\footnotesize Department of Materials Science and Engineering, University of Tennessee, Knoxville, TN 37996, United States}
\affiliation{\footnotesize Materials Science and Technology Division, Oak Ridge National Laboratory, Oak Ridge, TN 37831, United States}

\author{Kenji Watanabe}
\affiliation{\footnotesize National Institute for Material Science, 1-1 Namiki, Tsukuba 305-0044, Japan}

\author{Takashi Taniguchi}
\affiliation{\footnotesize National Institute for Material Science, 1-1 Namiki, Tsukuba 305-0044, Japan}

\author{Christian Sch{\"o}nenberger}
\email{Christian.Schoenenberger@unibas.ch}
\affiliation{\footnotesize Department of Physics, University of Basel, Klingelbergstrasse 82, CH-4056 Basel, Switzerland}
\affiliation{\footnotesize Swiss Nanoscience Institute, University of Basel,
Klingelbergstrasse 82, CH-4056 Basel, Switzerland}

\begin{abstract}
$\mathrm{WTe_2}$ is a material with rich topological properties: it is a 2D topological insulator as a monolayer and  a Weyl-semimetal and higher-order topological insulator (HOTI) in the  bulk form. Inducing superconductivity in topological materials is a way to obtain topological superconductivity, which lays at the foundation for many proposals of fault tolerant quantum computing. Here, we demonstrate the emergence of superconductivity at the interface between $\mathrm{WTe_2}$ and the normal metal palladium. The superconductivity has a critical temperature of about 1.2~\si{\kelvin}. By studying the superconductivity in perpendicular magnetic field, we obtain the coherence length and the London penetration depth. These parameters correspond to a low Fermi velocity and a high density of states at the Fermi level. This hints to a possible origin of superconductivity due to the formation of flat bands. Furthermore, the critical in-plane magnetic field exceeds the Pauli limit, suggesting a non-trivial nature of the superconducting state.

\end{abstract}
\maketitle

\section*{Introduction}
Topological materials attract a lot of attention in modern condensed matter physics. This interest stems from intriguing fundamental properties and great potential for practical applications. The especially interesting class of topological materials are topological superconductors, promising to revolutionize quantum computing due to the inherent error protection~\cite{quantcomp}. Topological superconductivity could be obtained by inducing superconductivity in topologically non-trivial system. There are several theoretical predictions of different topological superconducting states in Dirac and Weyl semimetal based systems, including Fulde--Ferrell–Larkin–Ovchinnikov superconductors~\cite{Cho,WheiHZ,Bednik}, the time-reversal invariant topological superconductor~\cite{Hosur}, chiral non-Abelian Majorana fermions\cite{Chan}, and flat band superconductivity~\cite{Tang}.

$\mathrm{WTe_2}$ is a layered transition-metal dichalcogenide with rich topological properties. As a bulk material it is a type-II Weyl semimetal with bulk Weyl nodes connected by Fermi arcs surface states~\cite{Soluyanov, Weyl_II}. Recently, it has been predicted to be a higher-order topological insulator with one-dimensional hinge states~\cite{HOTI}, experimental evidence of these states has been obtained~\cite{KononovNL,Choi,Huang}. In a single layer form, $\mathrm{WTe_2}$ is a two-dimensional topological insulator with helical edge states~\cite{Fei, Wu}. In addition to all these topological phases $\mathrm{WTe_2}$ has a tendency of becoming superconducting under different conditions: under pressure~\cite{presSC1,presSC2}, electron doping~\cite{dopSC} or electrostatic gating~\cite{gateSC1,gateSC2}. The combination of these properties makes $\mathrm{WTe_2}$ a particularly promising candidate for topological superconductivity.

In this manuscript we demonstrate the emergence of superconductivity at the interface between the normal metal palladium and few-layer thick $\mathrm{WTe_2}$. Studying the transport properties in magnetic field and at different temperatures we deduce the main parameters characterizing the superconducting state including the critical temperature, the coherence length and the London penetration depth. These parameters correspond to a low Fermi velocity and a high density of states at the Fermi level. This hints to a possible origin of superconductivity due to the formation of flat bands. Moreover, the measured in-plane critical field exceeds the Pauli limit, suggesting non-trivial superconducting pairing. The coexistence of the observed superconductivity with topological states in $\mathrm{WTe_2}$ makes it a promising platform for studying topological superconductivity and applications for quantum computing.

\section*{Experiment}
The single crystals of $\mathrm{WTe_2}$ were grown with a flux growth method~\cite{growth}. We obtained few-layer thick $\mathrm{WTe_2}$ flakes by mechanically exfoliating single crystals with an adhesive tape on oxidized Si substrate with 295~\si{\nano\meter} $\mathrm{SiO_2}$ layer. To avoid oxidation of $\mathrm{WTe_2}$, the exfoliation has been carried out in a glovebox with low oxygen content. We selected few-layer thick (5-12 single layers) stripe shaped flakes. Suitable flakes have been identified with an optical contrast method~\cite{Blake} and were picked up and transferred using the polycarbonate assisted pick-up technique~\cite{transfer} on the device chip that already contained prepatterned contacts. The contacts were defined before using standard e-beam lithography and metal deposition of 3~nm titanium and 12~nm palladium. In the final stack $\mathrm{WTe_2}$ is protected from oxidation by an hBN layer that covers the $\mathrm{WTe_2}$. All the measurements were performed in a dilution refrigerator with a base temperature of 60~\si{\milli\kelvin}. Similar superconducting properties have been observed in multiple devices, data presented in the manuscript were collected from 3 samples.

\section*{Results and discussion}
\begin{figure}[htb]
    \centering
    \includegraphics[width=\columnwidth]{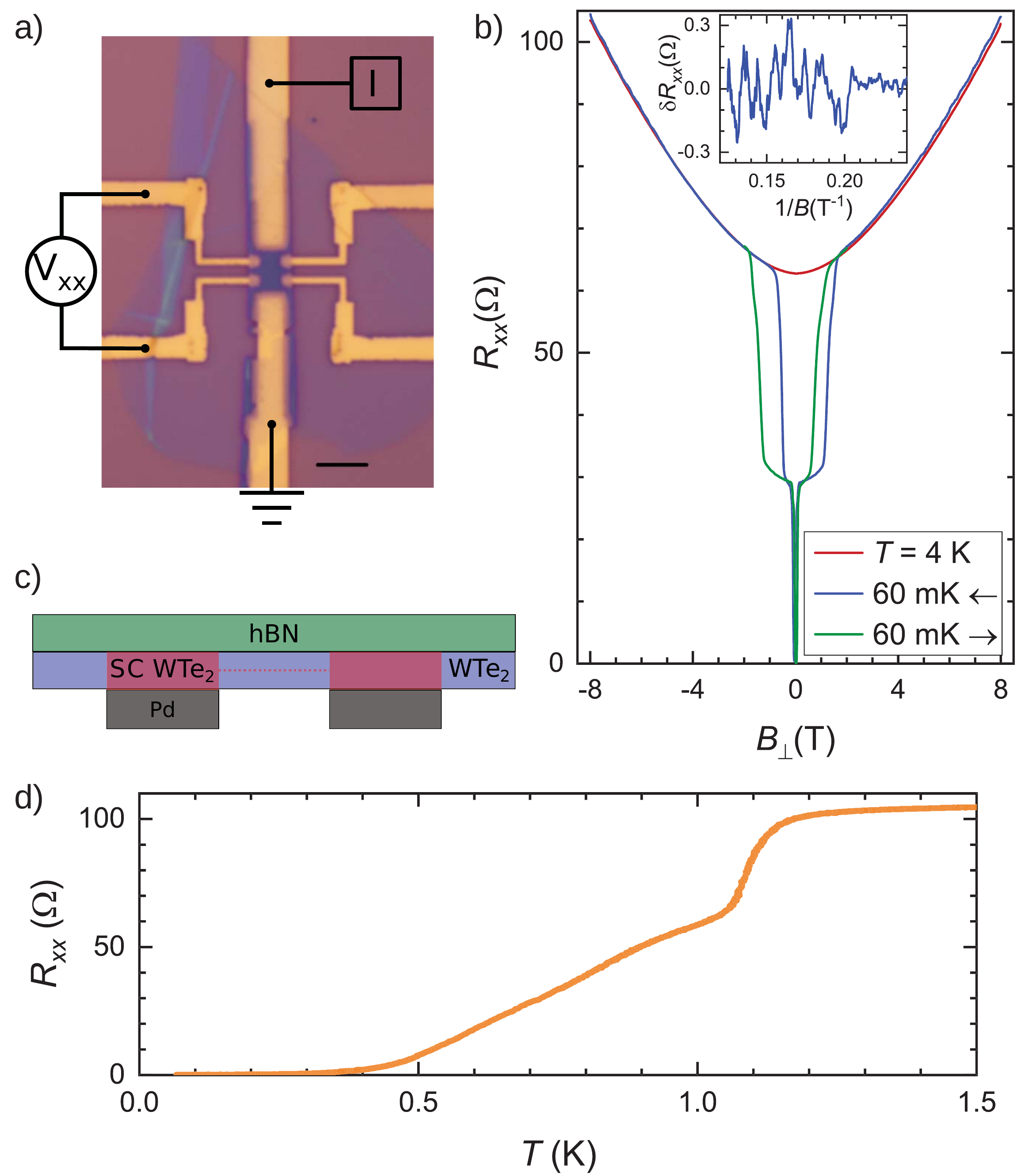}
    \caption{(\textbf{a}) Optical image of sample 1 (scale bar 5~\si{\micro\meter}) with a sketch of the measurement setup. (\textbf{b}) Longitudinal resistance $R_{xx}=V_{xx}/I$ as a function of perpendicular magnetic field $B_{\perp}$. At 4~\si{\kelvin} only a non-saturating magnetoresistance is seen, whereas at 60~\si{\milli\kelvin} the resistance shows additionally a transition to a smaller value in a magnetic field of $B_{\perp}\sim 1$\,T and a transition to zero-resistance due to emerging superconductivity for much lower $B_{\perp}$. The asymmetry in $R_{xx}(B)$ is likely connected with heating during magnetic field sweep, since it depends on the sweep direction and gets reduced with lower sweep rate. Inset: Shubnikov--de Haas oscillations highlighted by subtracting the overall quadratic trend from the 60~\si{\milli\kelvin} curve. (\textbf{c}) Cross-sectional view through the contact region: The region of $\mathrm{WTe_2}$ above the Pd leads turn into superconducting regions (red). These regions can be connected by the Josephson effect (red dashed line) if not too far apart. (\textbf{d}) Longitudinal resistance $R_{xx}$ as a function of temperature. Superconducting transition takes place in the range 1.05--1.2~\si{\kelvin}. The Josephson effect gradually develops at lower temperature achieving zero resistance state below 350~\si{\milli\kelvin}. Panels (a) and (c) reprinted with permission from A.~Kononov \textit{et al.}, Nano Lett. \textbf{20}, 4228 (2020) (\url{https://pubs.acs.org/doi/10.1021/acs.nanolett.0c00658}). Copyright 2020, American Chemical Society. Further permission related to the material excerpted should be directed to the ACS.}
    \label{R_xx}
\end{figure}

Fig.~\ref{R_xx}(a) shows an optical image of an encapsulated $\mathrm{WTe_2}$ crystal with a contact pattern that resembles a standard Hall-bar configuration. Note, that the visible Pd contacts are at the bottom, followed by a few layer $\mathrm{WTe_2}$ crystal with a rectangular shape and high-aspect ratio oriented vertically, followed by an hBN layer that has the weakest contrast in the image. The drawn electrical schematics corresponds to the measurement of the longitudinal  resistance $R_{xx}$ given by $V_{xx}/I$. 

Fig.~\ref{R_xx}(b) displays $R_{xx}$ as a function of perpendicular magnetic field $B_{\perp}$. At 4~\si{\kelvin} the resistance shows a non-saturating magnetoresistance characteristic for $\mathrm{WTe_2}$~\cite{WTe_qual}. The small thickness (7~layers) of our $\mathrm{WTe_2}$ crystal results in a relatively small magnetoresitance~\cite{Xiang}. Another evidence of the high quality of our samples is the presence of Shubnikov--de Haas oscillations at low temperature. The frequency of the oscillations $f_{1/B}\sim100$~\si{\tesla} corresponds to an electron density $n_{2D}=e/(\pi\hbar f_{1/B})\sim5\cdot10^{12}$~\si{\per\centi\meter\squared} and Fermi wavevector $k_F=\sqrt{\pi n_{2D}}\sim0.4$~\si{\per\nano\meter} (here the two electron pockets of $\mathrm{WTe_2}$ are taken into account). The oscillations visibility at around 5~\si{\tesla} suggest a mobility of at least 2000~\si{\centi\meter\squared\per\volt\per\second}, which yields an electron mean free path of $l_{mfp}=k_F\hbar\mu/e\sim50$~\si{\nano\meter}. 

At low temperature additional features develop in $R_{xx}(B_{\perp})$: at zero field the resistance goes to zero and in small fields it has an intermediate state between zero and high-temperature values. The asymmetry in $R_{xx}(B)$ is defined by the sweep direction and rate, thus it is likely connected with heating during the magnetic field sweep.
The intermediate resistance state in Fig.~\ref{R_xx}(b) is a result of the formation of a superconducting state in $\mathrm{WTe_2}$ above the Pd leads. Furthermore, these superconducting regions could be connected by the Josephson effect, as illustrated in Fig.~\ref{R_xx}(c), leading to a zero longitudinal resistance. The zero resistance state appears only for smaller distances between the contacts, excluding intrinsic superconductivity in our $\mathrm{WTe_2}$ samples. This explanation is further supported by $R_{xx}(T)$ dependence in Fig.~\ref{R_xx}(d). With decreasing temperature the first superconducting transition takes place in the range 1.05--1.2~\si{\kelvin}, followed by the gradual developing of the Josephson effect at lower temperature achieving zero resistance below 350~\si{\milli\kelvin}.

\begin{figure}[htb]
    \centering
    \includegraphics[width=\columnwidth]{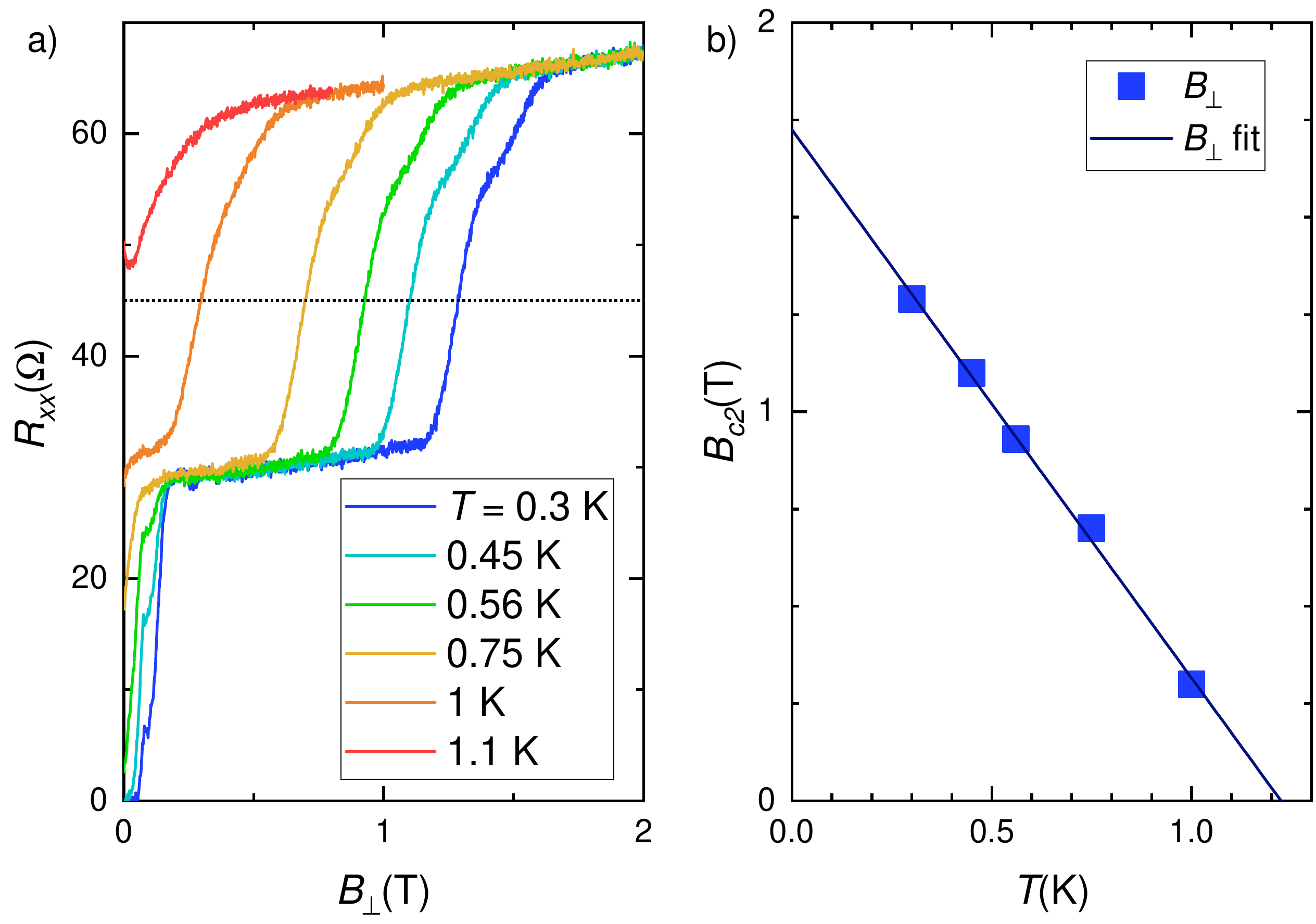}
    \caption{(\textbf{a}) Longitudinal resistance as a function of perpendicular magnetic field $B_{\perp}$ at different temperatures. The dotted line indicates the resistance value used to determine $B_{c2}(T)$. The increase in the resistance near $B_{\perp}=0$ is connected with heating of the sample during field sweeps.  (\textbf{b}) Critical magnetic field $B_{c2}$ as function of temperature extracted  from (a) and a $B_{c2}\propto(1-T/T_c)$ fit of the data. Panel (a) reprinted with permission from A.~Kononov \textit{et al.}, Nano Lett. \textbf{20}, 4228 (2020) (\url{https://pubs.acs.org/doi/10.1021/acs.nanolett.0c00658}). Copyright 2020, American Chemical Society. Further permission related to the material excerpted should be directed to the ACS.}
    \label{B_perp}
\end{figure}

To understand the properties of the superconducting state we studied the evolution of $R_{xx}(B_\perp)$ with increasing temperature, as shown in Fig.~\ref{B_perp}(a). Upon temperature increase both transitions in the resistance are shifting towards zero field. The zero resistance state connected to the Josephson coupling disappears first above 0.75~\si{\kelvin}, the second transition connected to the suppression of superconductivity by magnetic field $B_{c2}$ persists up to 1.1~\si{\kelvin}. We define $B_{c2}(T)$ as the magnetic field where the $R_{xx}(B_{\perp})$ crosses the fixed resistance value $R_{xx}=45$~\si{\ohm}, which approximately corresponds to half of the resistance step. Fig.~\ref{B_perp}(b) shows the extracted dependence of the critical magnetic field as a function of temperature ${T}$. The $B_{c2}(T)$ dependence is linear as expected for a 2D superconductor
\begin{equation}\label{2DGL}
B_{c2}(T)=\frac{\Phi_0}{2\pi\xi_{GL}^2}\left(1-\frac{T}{T_c}\right),\end{equation}
where $\Phi_0$ is the magnetic flux quantum, $\xi_{GL}$ is the Ginzburg-Landau coherence length at zero temperature, and $T_c$ is the critical temperature at zero magnetic field. Fitting the experimental data with equation~\ref{2DGL}, we obtain $T_c\sim1.2$~\si{\kelvin} and relatively short $\xi_{GL}\sim14$~\si{\nano\meter}.

Disorder can cause the reduction of the coherence length, but we don't think this is the case in our samples, since we have found $l_{mfp}>\xi$ and non-saturating magnetoresistance. In the clean limit at low temperatures the Ginzburg--Landau coherence length is similar to the Bardeen--Cooper--Schrieffer (BCS) coherence length $\xi_{GL}\sim\xi$. Knowing the coherence length and the critical temperature, we can estimate the Fermi velocity $v_F=\xi\pi\Delta/\hbar$, where we take for $\Delta(T_c)$ the BCS relation $\Delta\sim1.76k_BT_c$, yielding $v_F\sim1.2\cdot10^4$~\si{\meter\per\second}. The obtained small value of Fermi velocity could suggest superconductivity due to the formation of flat bands~\cite{Cao}.

\begin{figure}[htb]
    \centering
    \includegraphics[width=\columnwidth]{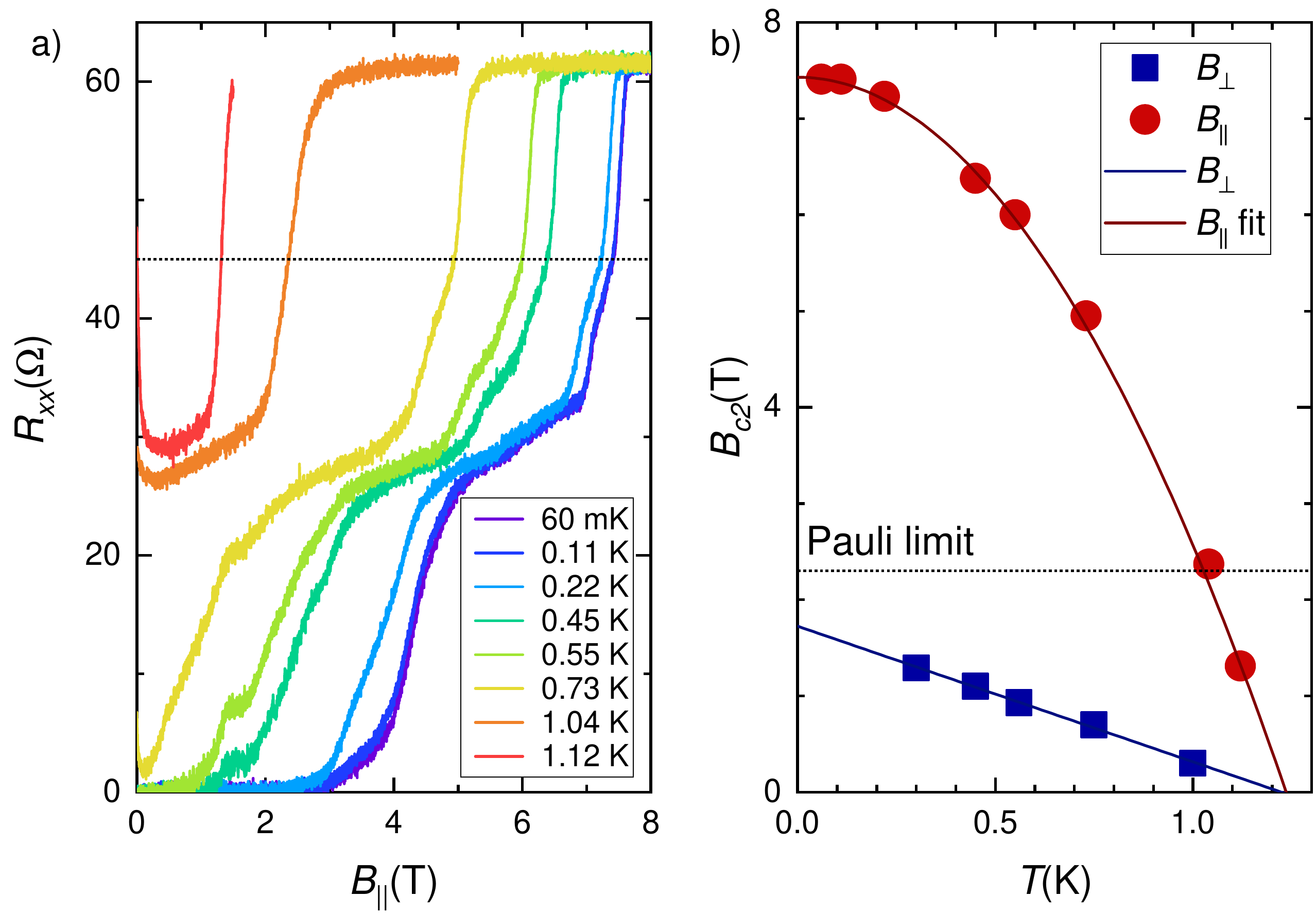}
    \caption{(\textbf{a}) Longitudinal resistance as a function of in-plane magnetic field at different temperatures. The dotted line indicates the resistance level used to determine $B_{c2}(T)$. The increase in the resistance near $B_{\parallel}=0$ is connected with heating of the sample during field sweeps. (\textbf{b}) Critical magnetic field as function of temperature extracted  from (a) and fit to the data. The dotted line indicates the maximum critical field given by the Pauli limit $B_P$, which we estimate to $2.3$~\si{\tesla}. The data for the critical perpendicular magnetic field is shown for comparison.}
    \label{B_par}
\end{figure}

We further investigate the superconducting properties by looking at the $R_{xx}$ dependence on the {\it in-plane} magnetic field $B_{\parallel}$, as shown in Fig.~\ref{B_par}(a). Compared with the perpendicular field, both changes in the resistance have shifted to higher magnetic fields. We extracted the critical field values as a function of temperature and plotted them in Fig.~\ref{B_par}(b). In this case, $B_{c2}(T)$ follows the known empirical law for superconductors $B_{c2}(T)=B_{c2}(0)\left[1-(T/T_c)^2\right]$~\cite{Tinkham}, as evident from the very good agreement between measured points and the fit. Both fits of the critical field as a function of $B_{\perp}$ and $B_{\parallel}$ converge to the same temperature $T_c\sim1.2$~\si{\kelvin}.

A notable feature of the parallel critical field is its large value, which exceeds the Pauli paramagnetic limit $B_P$. The latter is given by $B_P\sim1.76k_B T_c\sqrt{2}/g\mu_B\sim1.86T_c\sim2.3$~\si{\tesla}. This expression is based on the BCS theory for weak-coupling superconductors and a free electron $g$-factor of $g=2$~\cite{Saito}. This effect has also been observed in gated monolayer~\cite{gateSC1,gateSC2} and doped bulk $\mathrm{WTe_2}$~\cite{dopSC}, and ultrathin films of other materials~\cite{IsingSC1,IsingSC2}. Several mechanisms could be responsible for superconductivity exceeding the Pauli limit, including Ising-type superconductivity~\cite{Saito} or a diminishing of the effective $g$-factor due to strong spin-orbit coupling~\cite{Klemm}. Further studies are required to resolve this.

\begin{figure}[htb]
    \centering
    \includegraphics[width=\columnwidth]{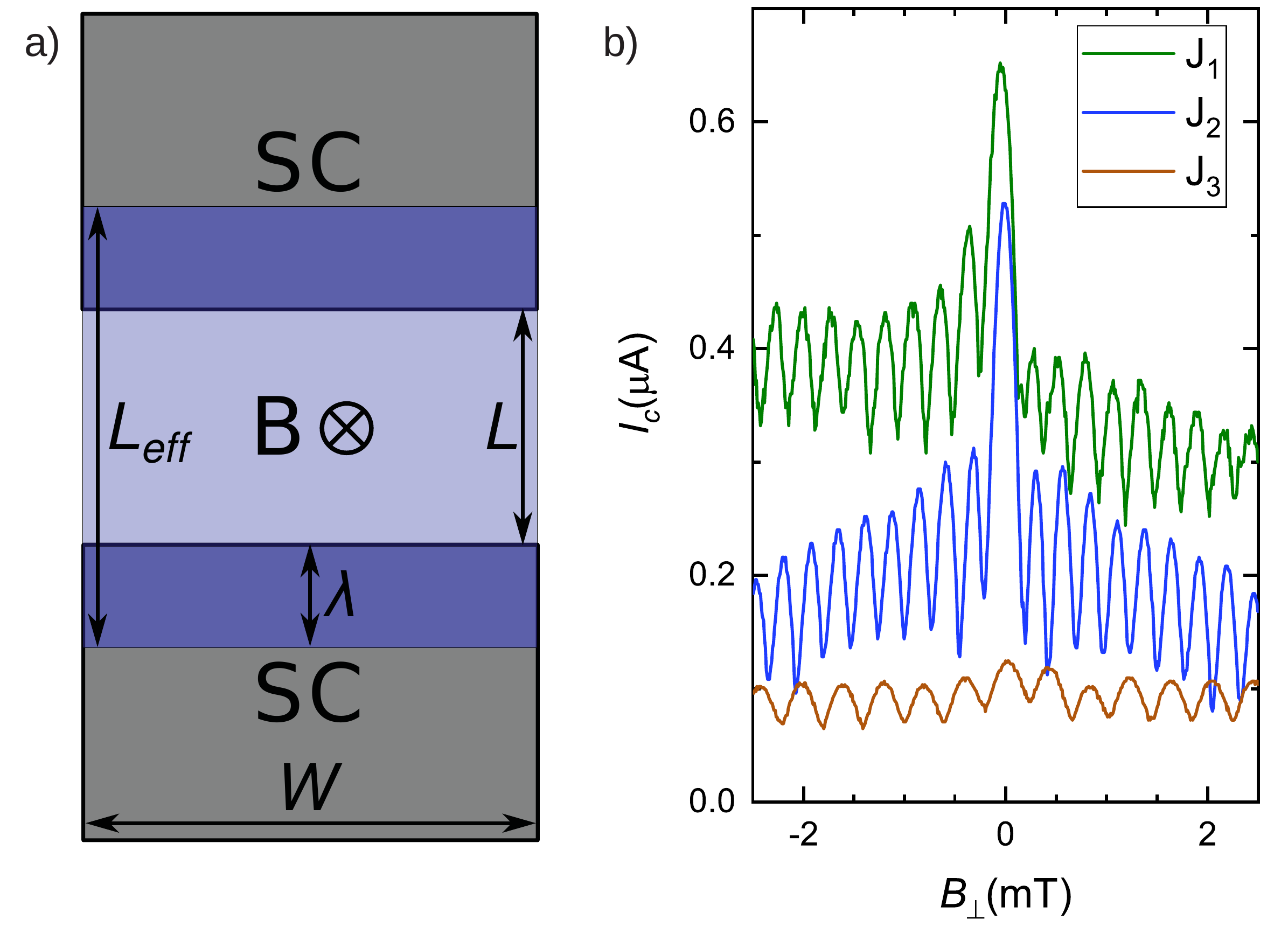}
    \caption{(\textbf{a}) Illustration of flux focusing in Josephson junction. The area of the junction, where the magnetic field is not screened and penetrating the junction,
is given by an effective area $W \times L_{eff}$, rather than the geometrical area $W \times L$, where $L$ is the junction length. Here, the effective length $L_{eff}$ equals $L+2\lambda$, where $\lambda$ is accounting for the penetration depth of the field into the superconductor. (\textbf{b}) Critical current as a function of magnetic field for three different Josephson junctions. Junctions $\mathrm{J_1}$ and $\mathrm{J_2}$ have $W=4.3~\si{\micro\meter}$ and length $L=1~\si{\micro\meter}$ and junction $\mathrm{J_3}$ has $W=4.2~\si{\micro\meter}$, $L=500~\si{\nano\meter}$. Data in (b) used with permission from A.~Kononov \textit{et al.}, Nano Lett. \textbf{20}, 4228 (2020) (\url{https://pubs.acs.org/doi/10.1021/acs.nanolett.0c00658}). Copyright 2020, American Chemical Society. Further permission related to the material excerpted should be directed to the ACS.}
    \label{I_osc}
\end{figure}

The London penetration depth $\lambda_L$ is another important characteristic of a superconductor. While RF-measurements are a common way of measuring the penetration depth~\cite{Gubin}, it can also be estimated by measuring the critical current of a Josephson junction in a magnetic field $B_{\perp}$. A Josephson junction placed in a perpendicular magnetic field demonstrates an oscillating critical current. One period of the oscillations corresponds to the magnetic flux quantum $\Phi_0$ through the effective area of the junction $S_{eff}=WL_{eff}=W(L+2\lambda)$, where $W$ and $L$ are the junction's width and length, respectively, and $\lambda$ is the magnetic field penetration depth~\cite{Tinkham}, see Fig~\ref{I_osc}(a). For a bulk superconductor $\lambda=\lambda_L$, but for a thin film superconductor with thickness $d$, the penetration depth is a function of the thickness $\lambda(d)=\lambda_L\coth{(d/\lambda_L)}$~\cite{Gubin}. In the limit of small thickness $d\ll\lambda_L$ the previous expression is equal to the Pearl's penetration depth $\lambda_P=\lambda_L^2/d$.

Fig.~\ref{I_osc}(b) demonstrates several examples of $I_c(B_\perp)$ dependencies for Josephson junctions in $\mathrm{WTe_2}$ where the superconducting regions on top of Pd play the role of superconducting contacts. These dependencies have a SQUID-like character due to hinge states~\cite{KononovNL} with a rapidly decaying Fraunhofer contribution due to the Fermi-arc surface states~\cite{Shevtsov2,Kononov} or the bulk conductivity. The SQUID-like oscillations with many visible periods allow to determine the period with a high precision. For junctions 1 and 2, with $L=1$~\si{\micro\meter} and $W=4.3$~\si{\micro\meter}, we obtain a period of ${\vartriangle}B=0.27$~\si{\milli\tesla}. This period corresponds to $L_{eff}=1.77$~\si{\micro\meter} and a penetration depth of $\lambda=380$~\si{\nano\meter}. For junction~3 ($L=500$~\si{\nano\meter}, $W=4.2$~\si{\micro\meter}) we obtain ${\vartriangle}B=0.41$~\si{\milli\tesla}, yielding $\lambda=350$~\si{\nano\meter}. The obtained penetration depth is much larger than the thickness of the $\mathrm{WTe_2}$ flakes $d\sim7$~\si{\nano\meter} (approximately 10-layers thick) so that the extracted penetration depth is given by Pearl's limit $\lambda=\lambda_P=\lambda_L^2/d$. Using this expression we estimate the London penetration depth to be $\lambda_L \sim 50$~\si{\nano\meter}. The ratio between the London penetration depth and the coherence length $\kappa=\lambda_L/\xi$ is $\kappa\sim3>1/\sqrt{2}$, suggesting type-II superconductivity~\cite{Tinkham}.

The obtained London penetration depth is comparable to typical values for metals and is surprisingly small considering the semi\-metalic nature of $\mathrm{WTe_2}$. This additionally speaks against the presence of disorder in our samples, since the penetration depth is expected to be higher in dirty superconductors~\cite{Tinkham}. An estimate of the superconducting electron density yields a quite high value $n_s=m/\mu_0\lambda_L^2e^2\sim3\cdot10^{21}$~\si{\per\cubic\centi\meter}, where $m\sim0.3m_e$ is the effective mass of the electrons in $\mathrm{WTe_2}$~\cite{Fatemi}. This value is higher than the typical carrier densities in $\mathrm{WTe_2}$ $n\sim10^{19}$~\si{\per\cubic\centi\meter}~\cite{Zhu} and corresponds to a density per single layer of $n_s^{1L}\sim2\cdot10^{14}$~\si{\per\square\centi\meter}, which is an order of magnitude higher than the electron density in monolayer $\mathrm{WTe_2}$ with gate induced superconductivity~\cite{gateSC1,gateSC2}, but comparable to the predicted optimal charge carrier density~\cite{SC_1L}. Furthermore, a large superconducting carrier density implies a high density of states at the Fermi level $g(E_F)\sim n_s/2\Delta\sim8\cdot10^{24}$~\si{\per\cubic\centi\meter\per\electronvolt}, which is a signature of flat bands.

The emergence of superconductivity at the interface of two non-superconducting materials is quite surprising, despite the fact that it has been observed previously in different Weyl and Dirac semimetals~\cite{Aggarwal,Wang_CdAs,Shvetsov,WZhu,XingY}. While the underlying mechanism for the superconductivity is unclear, in our case of $\mathrm{WTe_2}$/Pd interface several material specific reasons for the superconductivity could be proposed. First, the structural change at the interface could lead to the superconductivity similar to the pressure induced superconductivity in $\mathrm{WTe_2}$~\cite{presSC1, presSC2}. Second, electron doping from palladium~\cite{Shao} could create a superconducting state similar to what was seen in monolayer~\cite{gateSC1, gateSC2} or bulk doped $\mathrm{WTe_2}$~\cite{dopSC}. The latter seems to be the more probable explanation, since the in-plane critical field exceeds the Pauli limit, which has been observed in doped $\mathrm{WTe_2}$~\cite{gateSC1, gateSC2, dopSC}, but not in the pressure induced superconductivity~\cite{presSC1, presSC2}. Another possibility is interdiffusion of Pd and Te with a formation of superconducting $\mathrm{PdTe_2}$, which has been recently reported in samples with Pd deposited on $\mathrm{(Bi_{1-x}Sb_x)_2Te_3}$~\cite{Ando}. We think that this mechanism is unlikely in our case, since $\mathrm{WTe_2}$ and Pd are merely placed in contact by stacking and the samples were never heated above 180\si{\celsius}.

Even more intriguing is the possibility of the flat band superconductivity~\cite{Alidoust,Kauppila}, as suggested by the small Fermi velocity and high density of states at the Fermi level. Flat bands are ubiquitous in van der Waals (vdW) heterostructures. For example, high carrier-density combined with a low Fermi velocity has been observed close to van-Hove singularities in the band structure of superlattices formed in hBN-encapsulated graphene~\cite{Indolese}.  And the presence of the flat bands is known to stimulate superconductivity~\cite{Barash,Kopnin,Cao}.

Establishing the presence of the flat band superconductivity at $\mathrm{WTe_2}$/Pd interface and understanding the reasons for it will require further experiments, but some explanations could be outlined already. Flat band superconductivity can be formed as the result of a topological phase transition due to strain at the interface~\cite{Tang,Kauppila}. Furthermore, flat band superconductivity has been observed in vdW systems with moir\' e pattern~\cite{Cao}. This possibility is feasible since the  mismatch between lattice constant in Pd (0.389~\si{\nano\meter}~\cite{Davey}) and a-axis lattice constant in $\mathrm{WTe_2}$ (0.349\si{\nano\meter}~\cite{presSC2}) is only about 10\%. But we deem this  scenario unlikely, since moir\' e patterns strongly depend on the mutual orientation of the lattices, while we observe superconductivity in multiple samples without any intentional alignment of the lattices. 

An alternative explanation for the experimental data could be a multiband superconductivity in our samples. In this situation a sublinear dependence of $B_{c2}^{\perp}(T)$ and exceeding Pauli limit by $B_{c2}^{\parallel}$ could be expected in the dirty limit of the superconductivity~\cite{MultiB}, which seems not to be the case in our samples.

\section*{Conclusion}
We demonstrate the emergence of superconductivity at the interface between type-II Weyl semimetal $\mathrm{WTe_2}$ and normal metal palladium. Studying the transport properties in magnetic field and at different temperatures we deduce the key parameters that characterize the superconducting state, including the critical temperature $T_c$, the coherence length $\xi$ and the London penetration depth $\lambda_L$. The combined set of parameters hint to a possible origin of superconductivity being due to the formation of flat bands. Moreover, the measured in-plane critical field exceeds the Pauli limit, suggesting non-trivial superconducting pairing. The coexistence of superconductivity with topological states makes $\mathrm{WTe_2}$ a promising platform for topological superconductivity and applications for quantum computing.

\section*{Acknowledgments}
We thank A.~Baumgartner for helpful discussions.
A.K. was supported by the Georg H.~Endress foundation. This project has received further funding from the European Research Council (ERC) under the European Union’s Horizon 2020 research and innovation programme: grant agreement No 787414 TopSupra, by the Swiss National Science Foundation through the National Centre of Competence in Research Quantum Science and Technology (QSIT), and by the Swiss Nanoscience Institute (SNI).
K.W. and T.T. acknowledge support from the Elemental Strategy Initiative conducted by MEXT, Japan and the CREST (JPMJCR15F3), JST.
D.G.M. and J.Y. acknowledge support from the U.S. Department of Energy (U.S.-DOE), Office of Science - Basic Energy Sciences (BES), Materials Sciences and Engineering Division.
D.G.M. acknowledges support from the Gordon and Betty Moore Foundation’s EPiQS Initiative, Grant GBMF9069.

\section*{Data availability}
All data in this publication are available in numerical form in the Zenodo repository at \url{https://doi.org/10.5281/zenodo.3934679}~\cite{data}.

\bibliographystyle{abbrv}

\begin{thebibliography}{99}
\bibitem{quantcomp} S.~Nayak, S.H.~Simon, A.~Stern, M.~Freedman, and S.D.~Sarma, Non-Abelian anyons and topological quantum computation. \textit{Rev. Mod. Phys.} \textbf{80}, 1083 (2008).
\bibitem{Cho} G.Y.~Cho, J.H.~Bardarson, Y.M.~Lu, and J.E.~Moore, Superconductivity of doped Weyl semimetals: finite-momentum pairing and electronic analog of the 3He-A phase. \textit{Phys. Rev. B} \textbf{86}, 214514 (2012).
\bibitem{WheiHZ} H.Z.~Wei, S.P.~Chao, and V.~Aji, Odd-parity superconductivity in Weyl semimetals. \textit{Phys. Rev. B} \textbf{89}, 014506 (2014).
\bibitem{Bednik} G.~Bednik, A.A.~Zyuzin, and A.A.~Burkov, Superconductivity in Weyl metals. \textit{Phys. Rev. B} \textbf{92}, 035153 (2015).
\bibitem{Hosur} P.~Hosur, X.~Dai, Z.~Fang, and X.L.~Qi, Time-reversal-invariant topological superconductivity in doped Weyl semimetals. \textit{Phys. Rev. B} \textbf{90}, 045130 (2014).
\bibitem{Chan} C.~Chan and X.J.~Liu, Non-Abelian Majorana modes protected by an emergentsecond Chern number. \textit{Phys. Rev. Lett.} \textbf{118}, 207002 (2017).
\bibitem{Tang} E.~Tang and L.~Fu, Strain-induced partially flat band, helical snake states and interface superconductivity in topological crystalline insulators. \textit{Nat. Phys.} \textbf{10}, 964 (2014).
\bibitem{Soluyanov} A.A.~Soluyanov, D.~Gresch, Z.~Wang, Q.S.~Wu, M.~Troyer, X.~Dai, and B.A.~Bernevig, Type-II Weyl semimetals. \textit{Nature} \textbf{527}, 495–498 (2015).
\bibitem{Weyl_II} P.~Li, Y.~Wen, X.~He, Q.~Zhang, C.~Xia, Z.-M.~Yu, S.A.~Yang, Z.~Zhu, H.N.~Alshareef, and X.-X.~Zhang, Evidence for topological type-II Weyl semimetal $\mathrm{WTe_2}$. \textit{Nat. Commun.} \textbf{8}, 2150 (2017).
\bibitem{HOTI} Z.~Wang, B.J.~Wieder, J.~Li, B.~Yan, and B.A.~Bernevig, Higher-Order Topology, Monopole Nodal Lines, and the Origin of Large Fermi Arcs in Transition Metal Dichalcogenides $\mathrm{XTe_2}$ (X=Mo,W). \textit{Phys. Rev. Lett.} \textbf{123}, 186401 (2019).
\bibitem{KononovNL} A.~Kononov, G.~Abulizi, K.~Qu, J.~Yan, D.~Mandrus, K.~Watanabe, T.~Taniguchi, and C.~Sch{\"o}nenberger, One-Dimensional Edge Transport in Few-Layer $\mathrm{WTe_2}$. \textit{Nano Lett.} \textbf{20}, 4228 (2020).
\bibitem{Choi} Y.-B.~Choi, Y.~Xie, C.-Z.~Chen, J.~Park, S.-B.~Song, J.~Yoon, B.J.~Kim, T.~Taniguchi, K.~Watanabe, J.~Kim, K.C.~Fong, M.N.~Ali, K.T.~Law, and G.-H.~Lee, Evidence of Higher Order Topology in Multilayer $\mathrm{WTe_2}$ from Josephson Coupling through Anisotropic Hinge States. \textit{Nat. Mater.} \textbf{19}, 974 (2020).
\bibitem{Huang} C.~Huang, A.~Narayan, E.~Zhang, X.~Xie, L.~Ai, S.~Liu, C.~Yi, Y.~Shi, S.~Sanvito, and F.~Xiu, Edge superconductivity in Multilayer $\mathrm{WTe_2}$ Josephson junction. \textit{Natl. Sci. Rev.} \textbf{7}, 1468 (2020).
\bibitem{Fei} Z.~Fei, T.~Palomaki, S.~Wu, W.~Zhao, X.~Cai, B.~Sun, P.~Nguyen, J.~Finney, X.~Xu, and D.H.~Cobden, Edge conduction in monolayer $\mathrm{WTe_2}$. \textit{Nat. Phys.} \textbf{13}, 677 (2017).
\bibitem{Wu} S.~Wu, V.~Fatemi, Q.D.~Gibson, K.~Watanabe, T.~Taniguchi, R.J.~Cava, and P.~Jarillo-Herrero, Observation of the quantum spin Hall effect up to 100 Kelvin in a monolayer crystal. \textit{Science} \textbf{359}, 76 (2018).
\bibitem{presSC1} D.~Kang, Y.~Zhou, W.~Yi, C.~Yang, J.~Guo, Y.~Shi, S.~Zhang, Z.~Wang, C.~Zhang, S.~Jiang, A.~Li, K.~Yang, Q.~Wu, G.~Zhang, L.~Sun, and Z.~Zhao, Superconductivity emerging from a suppressed large magnetoresistant state in tungsten ditelluride. \textit{Nat. Commun.} \textbf{6}, 7804 (2015).
\bibitem{presSC2} X.-C.~Pan, X.~Chen, H.~Liu, Y.~Feng, Z.~Wei, Y.~Zhou, Z.~Chi, L.~Pi, F.~Yen, F.~Song, X.~Wan, Z.~Yang, B.~Wang, G.~Wang, and Y.~Zhang, Pressure-driven dome-shaped superconductivity and electronic structural evolution in tungsten ditelluride. \textit{Nat. Commun.} \textbf{6}, 7805 (2015).
\bibitem{dopSC} T.~Asaba, Y.~Wang, G.~Li, Z.~Xiang, C.~Tinsman, L.~Chen, S.~Zhou, S.~Zhao, D.~Laleyan, Y.~Li, Z.~Mi, and L.~Li, Magnetic Field Enhanced Superconductivity in Epitaxial Thin Film $\mathrm{WTe_2}$. \textit{Sci. Rep.} \textbf{8}, 6520 (2018).
\bibitem{gateSC1} E.~Sajadi, T.~Palomaki, Z.~Fei, W.~Zhao, P.~Bement, C.~Olsen, S.~Luescher, X.~Xu, J.A.~Folk, and D.H.~Cobden, Gate-induced superconductivity in a monolayer topological insulator. \textit{Science} \textbf{362}, 922 (2018).
\bibitem{gateSC2} V.~Fatemi, S.~Wu, Y.~Cao, L.~Bretheau, Q.D.~Gibson, K.~Watanabe, T.~Taniguchi, R.J.~Cava, P.~Jarillo-Herrero, Electrically tunable low-density superconductivity in a monolayer topological insulator. \textit{Science} \textbf{362}, 926 (2018).
\bibitem{growth} Y.~Zhao, H.~Liu, J.~Yan, W.~An, J.~Liu, X.~Zhang, H.~Wang, Y.~Liu, H.~Jiang, Q.~Li, Y.~Wang, X.-Z.~Li, D.~Mandrus, X.C.~Xie, M.~Pan, and J.~Wang, Anisotropic magnetotransport and exotic longitudinal linear magnetoresistance in $\mathrm{WTe_2}$ crystals. \textit{Phys. Rev. B} \textbf{92}, 041104(R) (2015).
\bibitem{Blake} P.~Blake, E.W.~Hill, A.H.~Castro Neto, K.S.~Novoselov, D.~Jiang, R.~Yang, T.J.~Booth, and A.K.~Geim, Making graphene visible. \textit{Appl. Phys. Lett.} \textbf{91}, 063124 (2007).
\bibitem{transfer} P.J.~Zomer, M.H.D.~Guimaraes, J.C.~Brant, N.~Tombros, and B.J.~van Wees, Fast pick up technique for high quality heterostructures of bilayer graphene and hexagonal boron nitride. \textit{Appl. Phys. Lett.} \textbf{105}, 013101 (2014).
\bibitem{WTe_qual} M.N.~Ali, L.~Schoop, J.~Xiong, S.~Flynn, Q.~Gibson, M.~Hirschberger, N.P.~Ong, and R.J.~Cava, Correlation of crystal quality and extreme magnetoresistance of $\mathrm{WTe_2}$. \textit{EPL} \textbf{110}, 67002 (2015).
\bibitem{Xiang} F.-X.~Xiang, A.~Srinivasan, Z.Z.~Du, O.~Klochan, S.-X.~Dou, A.R.~Hamilton, and X.-L.~Wang, Thickness-dependent electronic structure in $\mathrm{WTe_2}$ thin films. \textit{Phys. Rev. B} \textbf{98}, 035115 (2018).
\bibitem{Cao} Y.~Cao, V.~Fatemi, A.~Demir, S.~Fang, S.L.~Tomarken, J.Y.~Luo, J.D.~Sanchez-Yamagishi, K.~Watanabe, T.~Taniguchi, E.~Kaxiras, R.C.~Ashoori, and P.~Jarillo-Herrero, Correlated insulator behaviour at half-filling in magic-angle graphene superlattices. \textit{Nature (London)} \textbf{556}, 80 (2018).
\bibitem{Tinkham} M.~Tinkham, Introduction to superconductivity second edition. (McGraw-Hill, Inc., New York, 1996).
\bibitem{Saito} Y.~Saito, T.~Nojima, Y.~Iwasa, Highly crystalline 2D superconductors. \textit{Nat. Rev. Mater.} \textbf{2}, 16094 (2016).
\bibitem{IsingSC1} Y.~Liu, Y.~Xu, J.~Sun, C.~Liu, Y.~Liu, C.~Wang, Z.~Zhang, K.~Gu, Y.~Tang, C.~Ding, H.~Liu, H.~Yao, X.~Lin, L.~Wang, Q.-K.~Xue, and J.~Wang, Type-II Ising Superconductivity and Anomalous Metallic State in Macro-Size Ambient-Stable Ultrathin Crystalline Films. \textit{Nano Lett.} \textbf{20}, 5728 (2020).
\bibitem{IsingSC2} Y.~Liu, Z.~Wang, X.~Zhang, C.~Liu, Y.~Liu, Z.~Zhou, J.~Wang, Q.~Wang, Y.~Liu, C.~Xi, M.~Tian, H.~Liu, J.~Feng, X.C.~Xie, and J.~Wang, Interface-Induced Zeeman-Protected Superconductivity in Ultrathin Crystalline Lead Films. \textit{Phys. Rev. X} \textbf{8}, 021002 (2018).
\bibitem{Klemm} R.A.~Klemm, A.~Luther, M.R.~Beasley, Theory of the upper critical field in layered superconductors. \textit{Phys. Rev. B} \textbf{12}, 877 (1975).
\bibitem{Gubin} A.I.~Gubin, K.S.~Il’in, S.A.~Vitusevich, M.~Siegel, and N.~Klein, Dependence of magnetic penetration depth on the thickness of superconducting Nb thin films. \textit{Phys. Rev. B} \textbf{72}, 064503 (2005).
\bibitem{Shevtsov2} O.O.~Shvetsov, A.~Kononov, A.V.~Timonina, N.N.~Kolesnikov, and E.V.~Deviatov, Realization of a Double-Slit SQUID Geometry by Fermi Arc Surface States in a $\mathrm{WTe_2}$ Weyl Semimetal. \textit{JETP Lett.} \textbf{107}, 774 (2018).
\bibitem{Kononov} A.~Kononov, O.O.~Shvetsov, S.V.~Egorov, A.V.~Timonina, N.N.~Kolesnikov, and E.V.~Deviatov, Signature of Fermi arc surface states in Andreev reflection at the $\mathrm{WTe_2}$ Weyl semimetal surface. EPL \textbf{122}, 27004 (2018).
\bibitem{Fatemi} V.~Fatemi, Q.D.~Gibson, K.~Watanabe, T.~Taniguchi, R.J.~Cava, and P.~Jarillo-Herrero, Magnetoresistance and quantum oscillations of an electrostatically
tuned semimetal-to-metal transition in ultrathin  $\mathrm{WTe_2}$.  \textit{Phys. Rev. B} \textbf{95}, 041410(R) (2017).
\bibitem{Zhu} Z.~Zhu, X.~Lin, J.~Liu, B.~Fauqu\'e, Q.~Tao, C.~Yang, Y.~Shi, and K.~Behnia, Quantum oscillations, thermoelectric coefcients, and the Fermi surface of semimetallic $\mathrm{WTe_2}$. \textit{Phys. Rev. Lett.} \textbf{114}, 176601 (2015).
\bibitem{SC_1L} W.~Yang, C.-J.~Mo, S.-B.~Fu, Y.~Yang, F.-W.~Zheng, X.-H.~Wang, Y.-A.~Liu, N.~Hao, and P.~Zhang, Soft-mode-phonon-mediated unconventional superconductivity in monolayer 1T'-$\mathrm{WTe_2}$. \textit{Phys. Rev. Lett.} \textbf{125}, 237006 (2020).
\bibitem{Aggarwal} L.~Aggarwal, A.~Gaurav, G.S.~Thakur, Z.~Haque, A.K.~Ganguli, and G.~Sheet, Unconventional superconductivity at mesoscopic point contacts on the 3D Dirac semimetal $\mathrm{Cd_3As_2}$. \textit{Nat. Mat.} \textbf{15}, 32 (2016).
\bibitem{Wang_CdAs} H.~Wang, H.~Wang, H.~Liu, H.~Lu, W.~Yang, S.~Jia, X.-J.~Liu, X.C.~Xie, J.~Wei, and J.~Wang, Observation of superconductivity induced by a point contact on 3D Dirac semimetal $\mathrm{Cd_3As_2}$ crystals. \textit{Nat. Mat.} \textbf{15}, 38 (2016).
\bibitem{Shvetsov} O.O.~Shvetsov, V.D.~Esin, A.V.~Timonina, N.N.~Kolesnikov, and E.V.~Deviatov, Surface superconductivity in three-dimensional $\mathrm{Cd_3As_2}$ semimetal at the interface with a gold contact. \textit{Phys. Rev. B} \textbf{99}, 125305 (2019).
\bibitem{WZhu} W.~Zhu, X.~Hou, J.~Li, Y.~Huang, S.~Zhang, J.~He, D.~Chen, Y.~Wang, Q.~Dong, M.~Zhang, H.~Yang, Z.~Ren, J.~Hu, L.~Shan, G.~Chen, Interfacial superconductivity on the topological semimetal tungsten carbide induced by metal deposition. \textit{Adv. Mater.} \textbf{32}, 1907970 (2020).
\bibitem{XingY} Y.~Xing, Z.~Shao, J.~Ge, J.~Luo, J.~Wang, Z.~Zhu, J.~Liu, Y.~Wang, Z.~Zhao, J.~Yan, D.~Mandrus, B.~Yan, X.-J.~Liu, M.~Pan, J.~Wang, Surface superconductivity in the type II Weyl semimetal $\mathrm{TaIrTe_4}$. \textit{Nat. Sc. Rev.} \textbf{7}, 579, (2020).
\bibitem{Shao} B.~Shao, A.~Eich, C.~Sanders, A.S.~Ngankeu, M.~Bianchi, P.~Hofmann, A.A.~Khajetoorians, and T.O.~Wehling, Pseudodoping of a metallic two-dimensional material by the supporting substrate. \textit{Nat. Commun.} \textbf{10}, 180 (2019).
\bibitem{Ando} M.~Bai, F.~Yang, M.~Luysberg, J.~Feng, A.~Bliesener, G.~Lippertz, A.A.~Taskin, J.~Mayer, and Y.~Ando, Novel self-epitaxy for inducing superconductivity in the topological insulator $\mathrm{(Bi_{1-x}Sb_x)_2Te_3}$. \textit{Phys. Rev. Mat.} \textbf{4}, 094801 (2020).
\bibitem{Alidoust} M.~Alidoust, K.~Halterman, and A.A.~Zyuzin, Superconductivity in type-II Weyl semimetals. \textit{Phys. Rev. B} \textbf{95}, 155124 (2017).
\bibitem{Kauppila} V.J.~Kauppila, F.~Aikebaier, and T.T.~Heikkil{\"a}, Flat-band superconductivity in strained Dirac materials. \textit{Phys. Rev. B} \textbf{93}, 214505 (2016).
\bibitem{Indolese} D.I.~Indolese, R.~Delagrange, P.~Makk, J.R.~Wallbank, K.~Watanabe, T.~Taniguchi, and C.~Sch{\"o}nenberger, Signatures of van Hove singularities probed by the supercurrent in a graphene – hBN superlattice. \textit{Phys. Rev. Lett.}  \textbf{121}, 137701 (2018).
\bibitem{Barash} Yu.S.~Barash and P.I.~Nagornykh, Dispersionless modes and the superconductivity of ultrathin films. \textit{JETP Lett.} \textbf{83}, 376 (2006).
\bibitem{Kopnin} N.B.~Kopnin, T.T.~Heikkil{\"a}, and G.E.~Volovik, High-temperature surface superconductivity in topological flat-band systems. \textit{Phys. Rev. B} \textbf{83}, 220503(R) (2011).
\bibitem{Davey} W.P.~Davey, Precision measurements of the lattice constants of twelve common metals. \textit{Phys. Rev.} \textbf{25}, 753 (1925).
\bibitem{MultiB} X.~Xing, W.~Zhou, J.~Wang, Z.~Zhu, Y.~Zhang, N.~Zhou, B.~Qian, X.~Xu, and Z.~Shi, Two-band and pauli-limiting effects on the upper critical field of 112-type iron pnictide superconductors. \textit{Sci. Rep.} \textbf{7}, 45943 (2017).
\bibitem{data} A.~Kononov, Data for ``Superconductivity in type-II Weyl-semimetal WTe2 induced by a normal metal contact'', (Zenodo, 2020). \url{https://doi.org/10.5281/zenodo.3934679}
\end{thebibliography}

\end{document}